\documentclass[a4paper]{jpconf}
\usepackage{graphicx}

\def\a{\alpha}
\def\b{\beta}
\def\d{\delta}

\def\F{\Phi}
\def\g{\gamma}
\def\G{\Gamma}

\def\m{\mu}
\def\q{\psi}
\def\Q{\Psi}
\def\r{\rho}
\def\s{\sigma}
\def\S{\Sigma}
\def\t{\tau}

\def\callT{\mbox{$\mathcal{T}$}}

\def\bra{\langle}
\def\ket{\rangle}

\def\ra{\rightarrow}
\def\inf{\infty}

\newcommand{\be}{\begin{equation}}
\newcommand{\ee}{\end{equation}}
\newcommand{\bea}{\begin{eqnarray}}
\newcommand{\eea}{\end{eqnarray}}

\begin{document}
\title{Equilibrium and nonequilibrium 
many-body perturbation theory: a unified framework based on the 
Martin-Schwinger hierarchy}

\author{Robert van Leeuwen$^1$ and Gianluca Stefanucci$^2$}

\address{$^1$Department of Physics, Nanoscience Center, 
FIN 40014, University of Jyv\"askyl\"a, Jyv\"askyl\"a, Finland}
\address{$^2$Dipartimento di Fisica, Universit\`a di Roma Tor
Vergata, Via della Ricerca Scientifica 1, 00133 Rome, Italy}

\begin{abstract}
We present a unified framework for equilibrium and nonequilibrium many-body perturbation theory.
The most general nonequilibrium many-body theory valid for general initial states is based on a time-contour
originally introduced by Konstantinov and Perel'. The various other well-known formalisms of Keldysh, Matsubara and
the zero-temperature formalism are then derived as special cases that arise under different assumptions.
We further present a single simple proof of Wick's theorem that is at the same time valid in all these flavors
of many-body theory. It arises simply as a solution of the equations of the Martin-Schwinger hierarchy
for the noninteracting many-particle Green's function with appropriate boundary conditions. 
We further discuss a generalized Wick theorem for general initial states on the Keldysh contour and 
derive how the formalisms based on the Keldysh and Konstantinov-Perel'-contours are related for the
case of general initial states.   
\end{abstract}

\section{Introduction}

In many physical situations we are interested in knowing the 
expectation value of some observable quantity of a system in or out 
of equilibrium. For quantum systems of many identical and interacting 
particles
a very convenient mathematical object to extract  this information is the 
Green's function. Let $\hat{\rho}$ be the density matrix which describes 
the system at time, say, $t_{0}$ and $\hat{H}(t)$ be the Hamiltonian 
of the system for times $t>t_{0}$. The $n$-particle Green's function 
$G_{n}$ is defined according to
\be
G_{n}(1\ldots n;1'\ldots n')=\frac{1}{i^{n}}\Tr
\left[\hat{\rho}\;T\left\{
\hat{\q}_{H}(1)\ldots\hat{\q}_{H}(n)\hat{\q}_{H}^{\dag}(n')\ldots\hat{\q}_{H}(1')
\right\}\right].
\label{gn}
\ee
In this formula $1=(x_{1},t_{1})$, $2=(x_{2},t_{2})$, etc. are 
collective indices for the position-spin coordinates $x={\bf r},\s$ 
and time $t$, the symbol $\Tr$ denotes a trace over the Fock space, $T$ 
is the time-ordering operator and 
$\hat{\q}_{H}(j)=\hat{U}(t_{0},t_{j})\hat{\q}(j)\hat{U}(t_{j},t_{0})$ 
are field operators in the Heisenberg picture with respect to the 
Hamiltonian $\hat{H}$ (hence $\hat{U}$ is the evolution operator). 
The quantum average of a $n$-body operator can be calculated 
from the equal-time Green's function $G_{n}$.

The direct evaluation of $G_{n}$ from Eq. (\ref{gn}) is, in general, 
an impossible task. The first difficulty is brought by the Hamiltonian 
$\hat{H}=\hat{H}_{0}+\hat{H}_{\rm int}$ which is typically the sum of a one-body 
operator $\hat{H}_{0}$ and a $m$-body operator $\hat{H}_{\rm int}$ 
with $m\geq 2$. For $\hat{H}_{\rm int}\neq 0$
the field operator $\hat{\q}_{H}$ in the Heisenberg 
picture is a complicated object and must be approximated in some 
clever way. The second difficulty consists in 
taking the trace over the Fock space with a
density matrix $\hat{\r}$. The density matrix is a
self-adjoint, positive semi-definite operator with unit trace and, 
therefore, it can be written as
$\hat{\r}=e^{-\hat{X}}/\Tr[\hat{X}]$ where $\hat{X}$ is a self-adjoint 
operator. 
For instance for systems in equilibrium 
$\hat{X}=\b\hat{H}^{\rm M}$ with $\b$ the inverse temperature and 
$\hat{H}^{\rm M}=\hat{H}-\m\hat{N}$ the grand-canonical Hamiltonian. 
To make contact with this equilibrium situation we 
define $\hat{H}^{\rm M}=\hat{X}/\b$ so that 
\be
\hat{\r}=\frac{e^{-\b\hat{H}^{\rm M}}}{Z}
\ee
with $Z=\Tr[e^{-\b\hat{H}^{\rm M}}]$. In equilibrium $Z$ is the 
partition function. 
In order to specify the initial preparation of the system
we can assign either $\hat{\r}$ or $\hat{H}^{\rm M}$ since 
there is a one-to-one correspondence between the two.
If we now separate $\hat{H}^{\rm M}=\hat{H}^{\rm 
M}_{0}+\hat{H}^{\rm M}_{\rm int}$ into the sum of a one-body operator 
$\hat{H}^{\rm M}_{0}$ and a $m$-body operator $\hat{H}^{\rm M}_{\rm 
int}$ with $m\geq 2$ then the trace in Eq. (\ref{gn}) can easily be 
worked out for $\hat{H}^{\rm M}_{\rm int}=0$ whereas we have to use 
suitable approximation
schemes for 
$\hat{H}^{\rm M}_{\rm int}\neq 0$.

Different Many-Body Perturbation Theories (MBPT) have been put 
forward to overcome these difficulties. The most popular MBPT's are probably the 
zero-temperature (real-time) Green's Function Formalism 
(GFF) and the 
finite-temperature (imaginary-time) Matsubara GFF~\cite{fwbook}. These two formalisms are limited to  
equilibrium situations.
Systems driven out of equilibrium by an external field 
are usually studied within the (adiabatic real-time) Keldysh 
GFF~\cite{k.1965,d.1984}. The Keldysh GFF, however, neglects the effect of 
initial correlations which are relevant in the short-time dynamics of general
quantum systems, such as in transient dynamics in quantum
transport or in the study of atoms and molecules in external
laser fields. There exist two alternative GFF's to include initial 
correlations. The first is based on the idea of Konstantinov and 
Perel'~\cite{kp.1961} and consists in attaching the imaginary-time 
Matsubara track to the original Keldysh contour, see Refs. 
\cite{d.1984,w.1991,MozorovRoepke}. The second GFF does instead account for 
initial correlations through extra Feynman diagrams, the evaluation of 
which requires the knowledge of the reduced $n$-particle density 
matrices 
\be
\G_{n}(x_{1}\ldots x_{n};x'_{1}\ldots x'_{n})=\Tr[\hat{\r}\,
\hat{\q}^{\dag}(x'_{1})\ldots\hat{\q}^{\dag}(x'_{n})\hat{\q}(x_{n})\ldots\hat{\q}(x_{1})]
\label{gamman}
\ee
where the symbol $\Tr$ signifies a trace over the Fock space,
see Refs. \cite{h.1975,Bonitz:PRE95,Bonitz:JMP,bonitzbook,Garny,vls.2012}. 
These last two formalisms are both exact and hence equivalent.

In all the aforementioned GFF's the
dressed (interacting) $G_{n}$ is expanded in powers of the 
interaction Hamiltonian ($\hat{H}_{\rm int}$ and/or 
$\hat{H}_{\rm int}^{\rm M}$), leading to an expansion of  $G_{n}$ in terms
of the bare (noninteracting) Green's functions $G_{0,n}$. The 
appealing feature  
of any GFF is 
the possibility of reducing the $G_{0,n}$ to an (anti)symmetrized product of
$G_{0}\equiv G_{0,1}$ by means of Wick's theorem~\cite{w.1950}. Even though the 
 mathematical structure of all GFF's is identical, these formalisms 
are usually treated as independent probably due to the fact that the 
existing proofs of Wick's theorem are very much formalism-dependent. In this 
paper we show that Wick's theorem is the solution of a boundary 
problem for the Martin-Schwinger Hierarchy (MSH)~\cite{ms.1959} and that different GFF's correspond 
to different domains and parameters for the MSH~\cite{svlbook}. In this way we can 
easily explain the common mathematical structure of every GFF and see 
how, e.g., the Keldysh GFF reduces to the zero-temperature GFF in 
equilibrium or the Konstantinov-Perel' GFF reduces to the Keldysh 
GFF under the adiabatic assumption.
Our reformulation also allows us to prove a generalized Wick's 
theorem for interacting density matrices $\hat{\rho}$. This naturally 
leads to the 
diagrammatic expansion with extra Feynman diagrams  
previously mentioned. The generalized Wick expansion has a form identical
to that of a Laplace expansion for permanents/determinants
(for bosons/fermions). Consequently, the calculation of the
various prefactors is both explicit and greatly simplified. 
In this contribution we only state the generalized Wick's theorem and refer 
the reader to Refs. \cite{vls.2012,svlbook} for the proof. We will, 
however, discuss  the 
equivalence between the GFF based on the generalized Wick's theorem 
and the Konstantinov-Perel' GFF.

\section{General formula for the Green's function}

\begin{figure}[tbp]
\includegraphics[width=18pc]{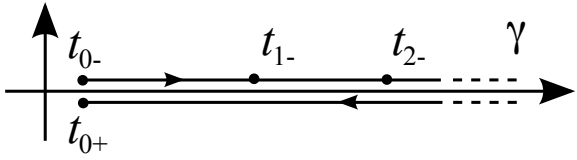}\hspace{2pc}%
\begin{minipage}[b]{18pc}
    \caption{\label{basiccontour}Contour $\g$ of Eq. (\ref{gn2}). The 
    contour consists of a forward branch going from $t_{0}$ to $\inf$ 
    (on this branch the points are denoted by $z=t_{-}$) and a backward branch 
    going from $\inf$ to $t_{0}$ (on this branch the points are 
    denoted by $z=t_{+}$).}
\end{minipage}
\end{figure}

The $n$-particle Green's function in Eq. (\ref{gn}) can also be 
written as
\be
G_{n}(1\ldots n;1'\ldots n')=
\frac{1}{i^{n}}\Tr\left[
\hat{\r}\;\callT\left\{
e^{-i\int_{\g}dz\hat{H}(z)}
\hat{\q}(1)\ldots\hat{\q}(n)\hat{\q}^{\dag}(n')\ldots\hat{\q}(1')
\right\}\right].
\label{gn2}
\ee
Let us explain this formula and discuss the equivalence with Eq. (\ref{gn}). 
In Eq. (\ref{gn2}) the integral is over the contour $\g$ of 
Fig. \ref{basiccontour} which goes from $t_{0}$ to $\inf$ and back to $t_{0}$ 
whereas $\callT$ is the contour ordering operator which rearranges 
operators with later contour arguments to the left. We denote by 
$z=t_{\pm}$ the points on $\g$ lying on the lower/upper branch at a 
distance $t$ from the origin and define the field operators with 
arguments on the contour as
\be
\hat{\q}(x,z)=\hat{\q}(x).
\ee
More generally every operator $\hat{O}(t)$ with a real-time argument 
can be converted into an operator $\hat{O}(z)$ with a contour-time 
argument according to the rule $\hat{O}(t_{+})=\hat{O}(t_{-})=\hat{O}(t)$. In 
particular $\hat{H}(t_{-})=\hat{H}(t_{+})=\hat{H}(t)$.
The reason to keep the 
contour argument in Eq. (\ref{gn2}) 
even for operators that do not have an explicit time 
dependence (like the field operators) stems from the need of 
specifying their position along the contour, thus rendering 
unambiguous the action of $\callT$. Once the operators are ordered we can 
omit the time arguments if there is no time dependence.         
For instance if $t_{1}<t_{2}$ then
\bea
\callT\left\{e^{-i\int_{\g}dz\hat{H}(z)}\hat{\q}(x_{1},t_{1-})
\hat{\q}^{\dag}(x_{2},t_{2-})\right\}&=&\pm
\hat{U}(t_{0},\inf)\hat{U}(\inf,t_{2})\hat{\q}^{\dag}(x_{2})
\hat{U}(t_{2},t_{1})\hat{\q}(x_{1})\hat{U}(t_{1},t_{0})
\nonumber \\
&=&T\left\{\hat{\q}_{H}(x_{1},t_{1})\hat{\q}^{\dag}_{H}(x_{2},t_{2})\right\},
\label{example}
\eea
where the $\pm$ sign in the first equality is for bosons/fermions. 
One can verify that Eq. (\ref{example}) is valid also for 
$t_{1}>t_{2}$.
This example can easily be generalized to many field operators. We 
conclude that Eq. (\ref{gn2}) is equivalent to Eq. (\ref{gn}) for 
contour arguments on the upper branch of $\g$. The $G_{n}$ in Eq. 
(\ref{gn2}) is, however, more general since the contour arguments can 
lie either on the upper or lower branch of $\g$. 
Quantities like photoemission currents, 
hyper-polarizabilities and more generally high-order response 
properties require the knowledge of this more general Green's 
function.

\begin{figure}[h]
\includegraphics[width=12pc]{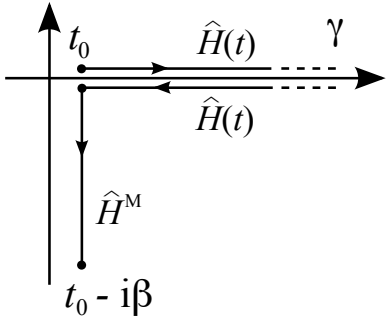}\hspace{4pc}%
\begin{minipage}[b]{21pc}
    \caption{\label{exactcontour}Contour $\g$ and Hamiltonian along 
    the contour $\g$ to get the exact (Konstantinov-Perel') Green's function from Eq. (\ref{gn3}).}
\end{minipage}
\end{figure}

The density matrix in Eq. (\ref{gn2}) can be incorporated into 
the contour ordering operator if we extend $\g$ as illustrated in 
Fig. \ref{exactcontour} and define the Hamiltonian with imaginary-time arguments as 
$\hat{H}(t_{0}-i\t)=\hat{H}^{\rm M}$. Since
\be
e^{-\b\hat{H}^{\rm M}}=e^{-i\int_{t_{0}}^{t_{0}-i\b}\hat{H}(z)}=
\callT\left\{e^{-i\int_{t_{0}}^{t_{0}-i\b}\hat{H}(z)}\right\}
\ee
we have
\be
G_{n}(1\ldots n;1'\ldots n')=
\frac{1}{i^{n}}\frac{\Tr\left[
\callT\left\{
e^{-i\int_{\g}dz\hat{H}(z)}
\hat{\q}(1)\ldots\hat{\q}(n)\hat{\q}^{\dag}(n')\ldots\hat{\q}(1')
\right\}\right]}{\Tr\left[
\callT\left\{
e^{-i\int_{\g}dz\hat{H}(z)}
\right\}\right]}
\label{gn3}
\ee
where in the denominator we took into account that $\callT\big\{
e^{-i\int_{t_{0-}}^{t_{0+}}dz\hat{H}(z)}\big\}=\hat{U}(t_{0},\inf)\hat{U}(\inf,t_{0})=1$.
Equation (\ref{gn3}) is, by construction, equivalent to Eq. 
(\ref{gn2}). It gives the {\em exact} Green's function provided that 
the integral is done along the contour $\g$ of Fig. \ref{exactcontour} and provided 
that the 
Hamiltonian changes along the contour as illustrated in the same 
figure. We now show that 
the Green's function of every GFF can be written as in Eq. (\ref{gn3}), 
the only difference being the shape of $\g$ and the Hamiltonian along 
$\g$.

We mentioned in the introduction that the calculation of the trace 
simplifies if the density 
matrix is of the form $\hat{\r}_{0}=e^{-\b\hat{H}_{0}^{\rm M}}/Z_{0}$, 
with $\hat{H}_{0}^{\rm M}$ a one-body operator and 
$Z_{0}=\Tr[e^{-\b\hat{H}_{0}^{\rm M}}]$. It is possible to turn a 
trace with $\hat{\rho}$ into a trace with $\hat{\rho}_{0}$ if the adiabatic 
assumption is fulfilled. According to the adiabatic assumption 
one can generate the density matrix $\hat{\r}$ with 
Hamiltonian $\hat{H}^{\rm M}=\hat{H}_{0}^{\rm M}+\hat{H}_{\rm 
int}^{\rm M}$ starting from the density matrix 
$\hat{\r}_{0}$ with Hamiltonian $\hat{H}_{0}^{\rm M}$ and then switching 
on $\hat{H}^{\rm M}_{\rm int}$ adiabatically, i.e.,
\be
\hat{\r}=\frac{e^{-\b\hat{H}^{\rm M}}}{Z}=
\hat{U}_{\eta}(t_{0},-\inf)\,\frac{e^{-\b\hat{H}_{0}^{\rm M}}}{Z_{0}}\,
\hat{U}_{\eta}(-\inf,t_{0})=\hat{U}_{\eta}(t_{0},-\inf)\,\hat{\r}_{0}\,
\hat{U}_{\eta}(-\inf,t_{0}),
\label{adass1}
\ee
where $\hat{U}_{\eta}$ is the real-time evolution operator with
Hamiltonian
\be
\hat{H}_{\eta}(t)=\hat{H}_{0}^{\rm M}+e^{-\eta|t-t_{0}|}\hat{H}^{\rm M}_{\rm int},
\nonumber
\ee
and $\eta$ is an infinitesimally small positive constant.
This Hamiltonian is equal to $\hat{H}_{0}^{\rm M}$ when 
$t\ra-\inf$ and is equal to the full interacting $\hat{H}^{\rm M}$ when $t=t_{0}$.
 In general the 
validity of the adiabatic assumption should be checked case 
by case. Under the adiabatic assumption we can rewrite Eq. 
(\ref{gn2}) as (omitting the arguments of $G_{n}$)
\be
G_{n}=
\frac{1}{i^{n}}\Tr\left[
\hat{\r}_{0}\;\hat{U}_{\eta}(-\inf,t_{0})\callT\left\{
e^{-i\int_{\g}dz\hat{H}(z)}
\hat{\q}(1)\ldots\hat{\q}(n)\hat{\q}^{\dag}(n')\ldots\hat{\q}(1')
\right\}\hat{U}_{\eta}(t_{0},-\inf)\right].
\label{gna}
\ee
\begin{figure}[tbp]
\includegraphics[width=18pc]{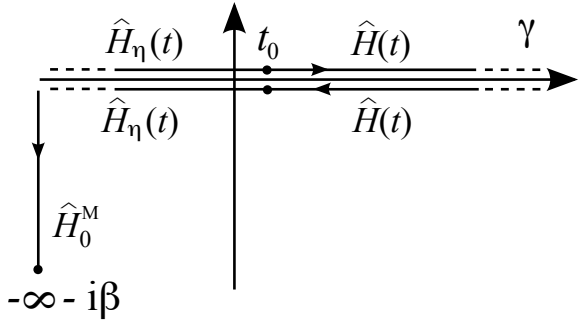}\hspace{4pc}%
\begin{minipage}[b]{16pc}
    \caption{\label{adiabaticcontour}Contour $\g$ and Hamiltonian along 
    the contour $\g$ to get the adiabatic (Keldysh) Green's function from Eq. (\ref{gn3}).}
\end{minipage}
\end{figure}
We now see that if we construct the contour 
$\g$ of Fig. \ref{adiabaticcontour} and let the Hamiltonian change along the contour as 
\bea
&&\hat{H}(t_{\pm})=\left\{\begin{array}{ll}
\hat{H}_{\eta}(t)=\hat{H}_{0}^{\rm M}+e^{-\eta|t-t_{0}|}\hat{H}^{\rm M}_{\rm int} & \quad\textrm{for 
$t<t_{0}$} \\ \\
\;\,\hat{H}(t)=\hat{H}_{0}(t)+\hat{H}_{\rm int}(t) &  \quad\textrm{for $t>t_{0}$} 
\end{array}\right.
\nonumber \\
\nonumber \\
&&\hat{H}(z\in \g^{\rm M})=\hat{H}_{0}^{\rm M}=\hat{H}_{0}-\m\hat{N},
\nonumber
\eea
then Eq. (\ref{gna}) takes the same form as Eq. (\ref{gn3}). We will 
refer to this way of calculating $G_{n}$ as the adiabatic formula.  
This is exactly the formula used by Keldysh in his original 
paper~\cite{k.1965}. The adiabatic formula is correct only provided that the 
adiabatic assumption is fulfilled.

We can derive yet another expression of $G_{n}$
for systems in equilibrium at zero temperature. In equilibrium
$\hat{H}^{\rm M}=\hat{H}-\m\hat{N}$ and therefore $\hat{H}_{0}^{\rm 
M}=\hat{H}_{0}-\m\hat{N}$ and $\hat{H}_{\rm int}^{\rm M}=\hat{H}_{\rm 
int}$. Assuming that $\hat{H}_{0}$ and $\hat{H}_{\rm int}$ commute 
with $\hat{N}$ the evolution operator $\hat{U}_{\eta}$ in Eq. 
(\ref{adass1}) can be calculated with Hamiltonian 
\be
\hat{H}_{\eta}(t)=\hat{H}_{0}+e^{-\eta|t-t_{0}|}\hat{H}_{\rm int}
\label{ztham}
\ee
since the addition of $-\m\hat{N}$ corresponds to multiplying 
$\hat{U}_{\eta}$ by a phase factor. In Eq. (\ref{adass1}) 
this phase factor cancels out since
$\hat{U}_{\eta}(-\inf,t_{0})=[\hat{U}_{\eta}(t_{0},-\inf)]^{\dag}$.
Furthermore, for any finite contour-times in $G_{n}$ we
can approximate the evolution operator $\hat{U}$ in the field 
operators $\hat{\q}_{H}$ with the evolution operator 
$U_{\eta}$ since we can always 
choose $\eta\ll 1/|t-t_{0}|$ and hence $\hat{H}_{\eta}\sim \hat{H}$. 
Thus Eq. (\ref{gna}) becomes
\be
G_{n}=
\frac{1}{i^{n}}\Tr\left[
\hat{\r}_{0}\;\callT\left\{
e^{-i\int_{\g}dz\hat{H}(z)}
\hat{\q}(1)\ldots\hat{\q}(n)\hat{\q}^{\dag}(n')\ldots\hat{\q}(1')
\right\}\right]
\label{gnzt}
\ee
where $\g$ is a contour that goes from $-\inf$ to $\inf$ and back 
to $-\inf$ and $\hat{H}(t_{\pm})=\hat{H}_{\eta}(t)$ is the 
Hamiltonian of Eq. (\ref{ztham}). Next we observe that the 
interacting  $\hat{\rho}$ can 
also be generated 
starting from $\hat{\r}_{0}$ and then propagating 
{\em backward} in time from $\inf$ to $t_{0}$ using the same evolution operator 
$\hat{U}_{\eta}$ since  
$\hat{H}_{\eta}(t_{0}-t)=\hat{H}_{\eta}(t_{0}+t)$. In other words
\be
\hat{\r}=\hat{U}_{\eta}(t_{0},\inf)\,\hat{\r}_{0}\,\hat{U}_{\eta}(\inf,t_{0}).
\nonumber
\ee
Comparing this equation with Eq. (\ref{adass1}) we conclude that
\be
\hat{\r}_{0}=\hat{U}_{\eta}(-\inf,\inf)\,\hat{\r}_{0}\,\hat{U}_{\eta}(\inf,-\inf).
\label{ro=ur0u}
\ee
If the ground state $|\F_{0}\ket$ of $\hat{H}_{0}-\m\hat{N}$ is nondegenerate then 
the zero-temperature $\hat{\r}_{0}=|\F_{0}\ket\bra\F_{0}|$ is a pure state and Eq. 
(\ref{ro=ur0u}) implies that
\be
\bra\F_{0}|\hat{U}_{\eta}(\inf,-\inf)=e^{i\a_{0}}\bra\F_{0}| .
\label{phaseapprx}
\ee
We will refer to the adiabatic assumption in 
combination with equilibrium at zero temperature and with the 
condition of no ground-state degeneracy as the 
{\em zero-temperature assumption}\index{zero-temperature assumption}. 
The zero-temperature assumption
can be used to manipulate Eq. (\ref{gnzt}) a bit more. 
We have
\be
\hat{\r}_{0}=|\F_{0}\ket\bra\F_{0}|=
\frac{|\F_{0}\ket\bra\F_{0}|\hat{U}_{\eta}(\inf,-\inf)}{\bra\F_{0}|\hat{U}_{\eta}(\inf,-\inf)|\F_{0}\ket}
\nonumber \\
=
\lim_{\b\ra\inf}
\frac{e^{-\b\hat{H}_{0}^{\rm M}}\hat{U}_{\eta}(\inf,-\inf)}{\Tr\left[
e^{-\b\hat{H}_{0}^{\rm M}}\hat{U}_{\eta}(\inf,-\inf)\right]}.
\nonumber
\ee
\begin{figure}[h]
\includegraphics[width=18pc]{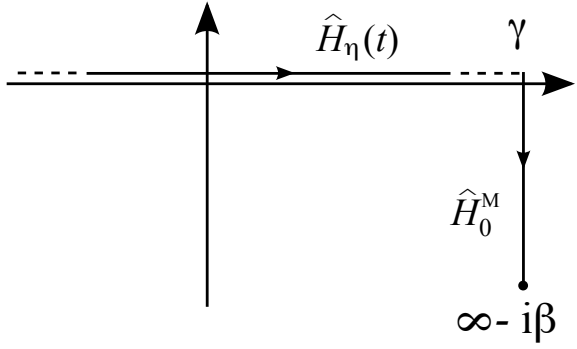}\hspace{4pc}%
\begin{minipage}[b]{16pc}
    \caption{\label{zerotemperaturecontour}Contour $\g$ and Hamiltonian along 
    the contour $\g$ to get the zero-temperature Green's function from Eq. (\ref{gn3}).}
\end{minipage}
\end{figure}
Inserting this result into Eq. (\ref{gnzt}) we find that the 
zero-temperature Green's function can again be written as in Eq. (\ref{gn3})
with the contour $\g$ that starts at $-\inf$, goes all the way to $\inf$ and then down to 
$\inf-i\b$, see Fig. \ref{zerotemperaturecontour}, and with the Hamiltonian $\hat{H}(z)$ that 
varies along the contour as illustrated in the same figure.
It is worth noticing that the contour $\g$ has the special property of having only a forward 
branch and that for the zero-temperature assumption to 
make sense the Hamiltonian of the system must be time independent. 
There is indeed no reason to expect that by switching on and  off 
the interaction the system goes back to the same state in the 
presence of external
driving fields.

To summarize the exact (Konstantinov-Perel'), adiabatic 
(Keldysh) and zero-temperature Green's 
functions have the same mathematical structure, given by Eq. (\ref{gn3}). What 
changes is the contour and the Hamiltonian along the contour.

\section{Wick's theorem and Many-Body Perturbation Theory}

To be concrete we specialize the discussion to interaction 
Hamiltonians $\hat{H}_{\rm int}$ and $\hat{H}_{\rm int}^{\rm M}$ 
which are two-body operators. Higher order $n$-body operators 
lead to more voluminous equations but do not rise conceptual 
complications. Thus we write
\be
\hat{H}_{\rm int}(z)=\frac{1}{2}\int dx_{1}dx_{2}\;
v(x_{1},x_{2};z)\hat{\q}^{\dag}(x_{1})\hat{\q}^{\dag}(x_{2})
\hat{\q}(x_{2})\hat{\q}(x_{1}).
\label{hint}
\ee
In the exact (Konstantinov-Perel') formula the interaction
$v(x_{1},x_{2};z)=v(x_{1},x_{2})$ is the interparticle interaction 
for $z$ on the horizontal branches whereas 
$v(x_{1},x_{2};z)$ depends on the initial preparation for $z$ on the vertical 
track. For 
instance in equilibrium $v(x_{1},x_{2};t_{0}-i\t)=v(x_{1},x_{2})$. On the 
other hand  in the adiabatic (Keldysh)
and zero-temperature formula $v(x_{1},x_{2};z)=e^{-\eta|t-t_{0}|}v(x_{1},x_{2})$
for $z$ on the horizontal branches whereas
$v(x_{1},x_{2};z)=0$ for $z$ on the vertical track.
Let us consider Eq. (\ref{gn3}) and write the exponential of $\hat{H}$ as the 
product of the exponentials of $\hat{H}_{0}$ and $\hat{H}_{\rm int}$:
\be
G_{n}(1\ldots n;1'\ldots n')=
\frac{1}{i^{n}}\frac{\Tr\left[
\callT\left\{
e^{-i\int_{\g}dz\hat{H}_{0}(z)}
e^{-i\int_{\g}dz\hat{H}_{\rm int}(z)}
\hat{\q}(1)\ldots\hat{\q}(n)\hat{\q}^{\dag}(n')\ldots\hat{\q}(1')
\right\}\right]}{\Tr\left[
\callT\left\{
e^{-i\int_{\g}dz\hat{H}_{0}(z)}e^{-i\int_{\g}dz\hat{H}_{\rm int}(z)}
\right\}\right]}.
\label{gn4}
\ee
The expansion in powers of $\hat{H}_{\rm int}$ leads to an expansion of 
$G_{n}$ in terms of noninteracting Green's functions $G_{0,n}$. 
The $G_{0,n}$ are obtained from Eq. (\ref{gn4}) by setting 
$\hat{H}_{\rm int}(z)=0$ for all $z\in\g$.
For instance for $n=1$ we get
\bea
G(a;b)=
\frac{\sum\limits^{\inf}_{k=0}\frac{1}{k!}\left(\frac{i}{2}\right)^{k}\int 
v(1;1')\ldots v(k;k')
G_{0,2k+1}(a,1,1',\ldots;b,1^{+},1'^{+},\ldots) }{
\sum\limits^{\inf}_{k=0}\frac{1}{k!}\left(\frac{i}{2}\right)^{k}\int 
v(1;1')\ldots v(k;k')
G_{0,2k}(1,1',\ldots;1^{+},1'^{+},\ldots)},
\label{GitG0n}
\eea
for $n=2$ we get
\be
G_{2}(a,b;c,d)=
\frac{
\sum\limits^{\inf}_{k=0}\frac{1}{k!}\left(\frac{i}{2}\right)^{k}\int
v(1;1')\ldots v(k;k')G_{0,2k+2}(a,b,1,1'\ldots;c,d,1^{+},1'^{+},\ldots)
}
{
\sum\limits^{\inf}_{k=0}\frac{1}{k!}\left(\frac{i}{2}\right)^{k}\int 
v(1;1')\ldots v(k;k')
G_{0,2k}(1,1',\ldots;1^{+},1'^{+},\ldots)
},
\ee
etc. In these equations $a=(x_{a},t_{a})$, $b=(x_{b},t_{b})$ are 
collective indices like $1,\; 2,\ldots$, the interaction 
$v(j;j')\equiv \d(z_{j},z'_{j})v(x_{j},x'_{j};z_{j})$ and the integrals 
are over $1,1',\ldots,k,k'$. 

The appealing feature of any GFF is the possibility of reducing the 
noninteracting $G_{0,n}$ to a (anti)symmetrized product for 
(fermions) bosons of one-particle Green's functions $G_{0}$. This 
reduction is called Wick's theorem. The existing proofs of Wick's 
theorem are rather laborious and differ
depending on whether one is working with the zero-temperature or 
Matsubara or Keldysh Green's functions. Below we give a 
simple and general proof of Wick's theorem which applies to all cases.

We consider a one-body Hamiltonian of the form
\be
\hat{H}_{0}(z)=\int dx\;\hat{\q}^{\dag}(x)h(x,z)\hat{\q}(x).
\label{h0}
\ee
The more general case of a nondiagonal $h(x,x',z)$ can be treated in 
a similar manner.
The Green's functions $G_{0,n}$ satisfy the noninteracting MSH
\be
\left[  i \frac{d}{dz_k} - h(k) \right] G_{0,n}
(1\ldots 
n;1'\ldots n')  
=\sum_{j=1}^n \, (\pm)^{k+j} \,\delta (k;j') \, G_{0,n-1} (1
\ldots \stackrel{\sqcap}{k}  \ldots n; 1' \ldots 
\stackrel{\sqcap}{j'} \ldots n' )  
\label{msoneG0}
\ee
\be
G_{0,n} (1\ldots n;1'\ldots n') \left[ - i 
\frac{\overleftarrow{d}}{dz_{k}'} - h(k') \right] 
=\sum_{j=1}^n \, (\pm)^{k+j} \,\delta (j;k') \,  
G_{0,n-1} (1 \ldots \stackrel{\sqcap}{j} \ldots n; 1' \ldots 
\stackrel{\sqcap}{k'} \ldots n' ) 
\label{mstwoG0}
\ee
where the hook over the arguments in $G_{0,n-1}$ means that those 
variables are missing. 
The MSH is a set of coupled differential equations to be solved 
on the contour $\g$ of the Green's function of interest (exact, 
adiabatic, or zero-temperature). In all cases 
from the definition Eq. 
(\ref{gn3}) it follows that the $G_{0,n}$ satisfy the 
Kubo-Martin-Schwinger (KMS) relations, i.e., the $G_{0,n}$ 
are (anti)periodic along the contour $\g$ with 
respect to all their contour arguments. Therefore we can calculate the $G_{0,n}$ 
by solving the MSH 
with KMS relations.  We now show that the solution is given by the 
Wick theorem 
\bea
G_{0,n} (1, \ldots, n; 1' ,\ldots ,n') = \left|
\begin{array}{ccc}
 G_0 (1;1') & \ldots  &G_0(1;n')    \\
 \vdots & & \vdots \\
 G_0 (n;1') & \ldots  & G_0(n;n')
\end{array}
\right|_{\pm}
\label{zeroGn}
\eea
where the symbol $|\ldots|_{\pm}$ signifies the permanent/determinant for 
the case of bosons/fermions and $G_{0}$ is the solution of Eqs. (\ref{msoneG0}) and 
(\ref{mstwoG0}) with $n=1$, i.e.,
\be
\left[  i \frac{d}{dz_1} - h(1) \right] G_{0}
(1;1')  =\d(1;1'),\quad\quad
G_{0} (1;1') \left[ - i 
\frac{\overleftarrow{d}}{dz_{1}'} - h(1') \right] 
=\d(1;1')
\label{mshg0}
\ee
with KMS boundary conditions. Expanding the permanent/determinant 
along row, say, $k$ we get
\be
G_{0,n} (1, \ldots, n; 1' ,\ldots ,n')=
\sum_{j=1}^{n}(\pm)^{k+j}G_{0}(k,j')
G_{0,n-1} (1
\ldots \stackrel{\sqcap}{k}  \ldots n; 1' \ldots 
\stackrel{\sqcap}{j'} \ldots n' )  
\ee
which is clearly a solution of Eq. (\ref{msoneG0}). Similarly, 
we can readily verify that  Eq. (\ref{zeroGn}) is also solution of Eq. 
(\ref{mstwoG0}) by expanding the permanent/determinant along column 
$k$. It remains to check that the $G_{0,n}$ in Eq. (\ref{zeroGn}) 
fulfills the KMS relations. The contour argument $z_{k}$ appears in 
all the $G_{0}$ of the $k$-th row of Eq. (\ref{zeroGn}) and nowhere 
else. Therefore when we move $z_{k}$ from the starting to the ending 
point of $\g$ all entries of row $k$ pick up a $(\pm)$ sign. Since 
the permanent/determinant of a matrix in which we multiply a row by 
$(\pm)$ is $(\pm)$ the permanent/determinant of the original matrix 
we conclude that $G_{0,n}$ is (anti)periodic with respect to the first 
$n$ contour arguments. With a similar reasoning one can prove that 
$G_{0,n}$ is (anti)periodic with respect to the last 
$n$ contour arguments. This concludes the proof.

The Wick theorem has been proven without any assumption on the shape 
of the contour and without any assumption on the form of the 
single-particle Hamiltonian $h(x,z)$ along the contour. Inserting 
Eq. (\ref{zeroGn}) into, e.g., Eq. (\ref{GitG0n}) we get the 
MBPT formula for the one-particle Green's 
function
\be
G(a;b)=\frac{\sum\limits^{\inf}_{k=0}\frac{1}{k!}\left(\frac{i}{2}\right)^{k}\!\int \!
v(1;1')\,..\, v(k;k')\!\left|\!\!\!
\begin{array}{cccc}
G_{0}(a;b) & G_{0}(a;1^{+}) & \ldots & G_{0}(a;k'^{+}) \\
G_{0}(1;b) & G_{0}(1;1^{+}) & \ldots & G_{0}(1;k'^{+}) \\
\vdots & \vdots & \ddots & \vdots \\
G_{0}(k';b) & G_{0}(k';1^{+}) & \ldots & G_{0}(k';k'^{+})
\end{array}\!\!\!\right|_{\pm}}
{
\sum\limits^{\inf}_{k=0}\frac{1}{k!}\left(\frac{i}{2}\right)^{k}\!\int \!
v(1;1')\,..\, v(k;k')\!
\left|\!\!\!\begin{array}{cccc}
G_{0}(1;1^{+}) & G_{0}(1;1'^{+}) & \ldots & 
G_{0}(1;k'^{+}) \\
G_{0}(1';1^{+}) & G_{0}(1';1'^{+}) & \ldots & 
G_{0}(1';k'^{+}) \\
\vdots & \vdots & \ddots & \vdots \\
G_{0}(k';1^{+}) & G_{0}(k';1'^{+}) & \ldots & 
G_{0}(k';k'^{+}) 
\end{array}\!\!\!\right|_{\pm}\!\!
}
\label{magicg}
\ee
which is an exact expansion of the interacting $G$ in terms of the 
noninteracting $G_{0}$. The MBPT for higher order Green's functions 
can be derived similarly. In the next Section we discuss how the 
variuos GFF's follow from Eq. (\ref{magicg}).

\section{Matsubara, Keldysh and zero-temperature formalisms}

\begin{figure}[h]
\includegraphics[width=18pc]{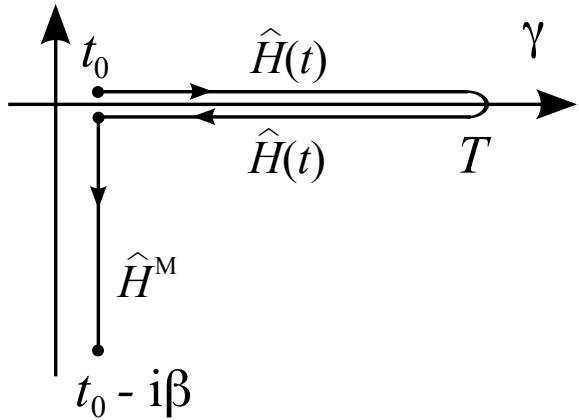}\hspace{4pc}%
\begin{minipage}[b]{16pc}
    \caption{\label{shrunkencontour}The shrunken contour $\g$ that 
    can be used in Eq. (\ref{magicg})  to obtain the Green's 
    function with real-time smaller than $T$.}
\end{minipage}
\end{figure}

In the Konstantinov-Perel' formalism the Green's function 
$G$ is given by Eq. (\ref{magicg}) where the 
$z$-integrals run on the contour of Fig. \ref{exactcontour}. 
It is worth stressing that if the times $t_{a}$ and $t_{b}$ in 
$G(a;b)$ are smaller than a 
maximum time $T$ then it is sufficient to perform the $z$ integrals 
over a shrunken contour like the one 
illustrated in Fig. \ref{shrunkencontour}. 
This is a direct consequence of the fact that if the contour is 
longer than $T$ then the terms with integrals after $T$  
cancel off~\cite{svlbook}.

The Matsubara GFF is used to calculate Green's function with 
imaginary times and it is typically applied to systems in equilibrium
at finite temperature. For this reason the Matsubara GFF is also called 
the ``finite-temperature formalism''. To calculate $G$ with 
imaginary-time arguments we can choose $T=t_{0}$ in Fig. \ref{shrunkencontour}
and hence shrink the horizontal branches to a point 
leaving only the vertical track. 
Therefore the Matsubara GFF consists of expanding the Green's 
function as in Eq. (\ref{magicg}) with the $z$-integrals 
restricted to the vertical track. It is important to realize that 
no assumptions, like the adiabatic or the zero-temperature 
assumption, are made in this formalism. The Matsubara GFF is 
exact but limited to initial (or equilibrium) averages. Equivalently 
we can say that the Matsubara $G$ is the same as the 
Konstantinov-Perel' $G$ on the vertical track.

The formalism originally used by Keldysh was based on the adiabatic 
assumption. The Keldysh Green's functions are again given by 
Eq. (\ref{magicg}) but the $z$-integrals are done over the contour of 
Fig. \ref{adiabaticcontour} and the Hamiltonian changes along the contour as illustrated 
in the same figure. 
The important simplification of the Keldysh GFF 
is that the interaction $v$ is zero on the vertical track.
Consequently in Eq. (\ref{magicg}) we can restrict the $z$-integrals 
to the horizontal branches. Like the  Konstantinov-Perel' formalism,  the 
Keldysh GFF can be used to deal with nonequilibrium situations in which 
the external perturbing fields are switched on after time $t_0$. In 
the special case of no external fields we can 
calculate interacting equilibrium Green's functions at 
any finite temperature with real-time arguments. 

The zero-temperature formalism relies on the
zero-temperature assumption. As we already discussed this 
assumption makes sense only in the absence of 
external fields.  The 
corresponding zero-temperature Green's function
is given by Eq. (\ref{magicg}) in which the $z$-integrals 
are done over the contour of Fig. \ref{zerotemperaturecontour} and the Hamiltonian changes along the contour as illustrated 
in the same figure. Like in the Keldysh GFF  the 
interaction $v$ vanishes along the 
vertical track and hence the $z$-integrals 
can be restricted to a contour that goes from $-\inf$ 
to $\inf$. The contour ordering operator
is then the same as the standard time-ordering operator. 
For this reason the 
zero-temperature Green's function is also called time-ordered 
Green's function. The zero-temperature GFF allows 
us to calculate the interacting $G$ in equilibrium at zero temperature 
with real-time arguments. It cannot, however, be used to study systems 
out of equilibrium and/or at finite temperature. In some cases, 
however, the zero-temperature formalism is used also at finite 
temperatures (finite $\b$) as the finite temperature corrections are 
small.
This approximated formalism is sometimes referred to as the {\em 
real-time finite temperature formalism}~\cite{fwbook}.
We emphasize that in the real-time finite-temperature formalism (like in the 
Keldysh formalism)  the temperature 
enters in Eq. (\ref{magicg}) only through $G_{0}$ which satisfies the KMS 
relations. In the  
Konstantinov-Perel' formalism, on the other 
hand, the temperature enters through $G_{0}$ and through the contour 
integrals since the interaction is nonvanishing along the vertical 
track.

\section{Generalized Wick's theorem}

The MBPT of the exact GFF requires the knowledge of the 
operator $\hat{H}^{\rm M}$ on the vertical track. In many physical 
situations, however, it is easier to specify the initial state (or initial density 
matrix) instead of $\hat{H}^{\rm M}$. In these cases 
the preliminary step to apply MBPT consists in obtaining $\hat{H}^{\rm M}$ from 
$\hat{\r}$, something that can be rather awkward. For instance if 
$\hat{\r}=|\Q\ket\bra\Q|$ is a pure state then $\hat{H}^{\rm M}$ is 
an operator with $|\Q\ket$ as the ground state. The 
existence of a generalized Wick's theorem that uses directly 
$\hat{\r}$  would be of very valuable. In this Section we will show 
how to construct such a generalized framework.

We consider again Eq. (\ref{gn2}) but this time we do not incorporate 
$\hat{\r}$ in the contour ordering. In Eq. (\ref{gn2}) the contour 
$\g$ is that of Fig. \ref{basiccontour} and the Hamiltonian is the physical 
Hamiltonian which, for simplicity, we take as the sum of $\hat{H}_{0}$ 
in Eq. (\ref{h0}) and $\hat{H}_{\rm int}$ in Eq. (\ref{hint}). We 
write the exponential in Eq. (\ref{gn2}) as the product of two 
exponentials, one containing $\hat{H}_{0}$  and the other containing 
$\hat{H}_{\rm int}$, like we did in Eq. (\ref{gn4}). The subsequent 
expansion of $G$ in powers of $\hat{H}_{\rm int}$ leads to the 
expansion 
\bea
G(a;b)=
\sum\limits^{\inf}_{k=0}\frac{1}{k!}\left(\frac{i}{2}\right)^{k}\int 
v(1;1')\ldots v(k;k')
g_{2k+1}(a,1,1',\ldots;b,1^{+},1'^{+},\ldots) 
\label{genexp}
\eea
and similarly for higher order Green's function. In Eq. 
(\ref{genexp}) the Green's functions $g_{n}$ are noninteracting 
Green's functions averaged with an arbitrary density matrix $\hat{\r}$
\be
g_{n}(1\ldots n;1'\ldots n')=
\frac{1}{i^{n}}\Tr\left[
\hat{\r}\;\callT\left\{
e^{-i\int_{\g}dz\hat{H}_{0}(z)}
\hat{\q}(1)\ldots\hat{\q}(n)\hat{\q}^{\dag}(n')\ldots\hat{\q}(1')
\right\}\right].
\label{gengn}
\ee
We will now prove a generalized Wick theorem to write these $g_{n}$ 
in terms of the one-particle Green's function $g\equiv g_{1}$ and 
the $n$-particle reduced density matrices $\G_{n}$ defined in Eq. 
(\ref{gamman}).
 
The Green's functions $g_{n}$ satisfy the noninteracting MSH on the 
contour of Fig. \ref{basiccontour}. The problem in solving the MSH to obtain the 
$g_{n}$'s is that we cannot use the KMS relations as boundary 
conditions. Indeed it is easy to verify that the $g_{n}$ are not 
(anti)periodic along the contour.
A convenient choice of boundary conditions follows 
directly from the definition of $g_{n}$ and reads
\be
(\pm i)^{n}\lim_{z_{k},z'_{j}\ra t_{0-}}
g_{n}(1\ldots n;1',\ldots n')=\G_{n}(x_{1}\ldots x_{n};x'_{1}\ldots x'_{n})
\label{bcgenw}
\ee
where the limit is taken with the order 
$z_{1}<\ldots<z_{n}<z'_{n}<\ldots<z'_{1}$ of the contour arguments. The permanent/determinant
\be
g_{n}(1\ldots n;1'\ldots n')=
\left|
\begin{array}{ccc}
g(1;1') & É & g(1;n') \\
\vdots & \vdots & \vdots \\
g(n;1') & É & g(n;n')
\end{array}
\right|_\pm\equiv |g|_n(1\ldots n;1'\ldots n')
\label{pdet}
\ee
is a solution of the MSH but, in general, with the wrong boundary conditions.
In Eq. (\ref{pdet}) the symbol $|\ldots|_{\pm}$ signifies the 
permanent/determinant of the matrix inside the vertical bars.
The particular solution must be supplied with the solution of the 
homogeneous equations
\be
\left[  i \frac{d}{dz_k} - h(k) \right] \tilde{g}_{n}
(1\ldots 
n;1'\ldots n')  
=0
\label{mshhom}
\ee
\be
\tilde{g}_{n} (1\ldots n;1'\ldots n') \left[ - i 
\frac{\overleftarrow{d}}{dz_{k}'} - h(k') \right] 
=0
\label{mshhom2}
\ee
to satisfy the correct boundary conditions. We observe that 
$\tilde{g}_{n}$ is not discontinuous when its contour arguments cross 
each other since in the right hand side of Eqs. (\ref{mshhom}) and (\ref{mshhom2}) there is no 
$\d$-function. Consequently the equal-time limit of $\tilde{g}_{n}$ 
is independent of the order of the contour arguments. 

Let us start by showing how to solve the MSH for $g_{2}$. We write 
$g_{2}=|g|_{2}+\tilde{g}_{2}$ where $g$ satisfies the first equation of 
the MSH with boundary conditions 
\be
(\pm i)\lim_{z_{1},z'_{1}\ra 
t_{0-}}g(1;1')=\G_{1}(x_{1};x'_{1})\equiv \G(x_{1};x'_{1}).
\label{g_0}
\ee
The boundary conditions for $\tilde{g}_{2}$ follow directly from Eq. 
(\ref{bcgenw}) and read
\be
(\pm i)^2\lim_{z_k,z'_j\to t_{0-}}\tilde{g}_2
=(\pm i)^2\lim_{z_k,z'_j\to t_{0-}}(g_2-|g|_2)=
\Gamma_2-|\Gamma|_2\equiv C_2.
\label{bcfg2}
\ee
Next we consider the spectral function on the contour
\be
A(1;1')=i\left[g^{>}(1;1')-g^{<}(1;1')\right].
\label{spectral}
\ee
This function takes the same value for $z_{1}=t_{1\pm}$ and 
$z'_{1}=t'_{1\pm}$ and satisfies the equations
\be
\left[i\frac{d}{dz_1}-h(1)\right]A(1;1')=A(1;1')\left[-i\frac{\overleftarrow{d}}{dz'_1}-h(1')\right]=0.
\label{spectral_mot}
\ee
Furthermore, due to the (anti)commutation rules of the field operators
\be
A(x_{1},z;x'_{1},z)=\d(x_{1}-x'_{1}).
\ee
Therefore 
\bea
\tilde{g}_2(1,2;1',2')&=&\int d\bar{x}_1d\bar{x}_2d\bar{x}'_1d\bar{x}'_2
A(1;\bar{x}_1,t_{0-})A(2;\bar{x}_2,t_{0-})C_2(\bar{x}_1,\bar{x}_2;\bar{x}'_1,\bar{x}'_2)
\nonumber \\
&\times&
A(\bar{x}'_1,t_{0-};1')A(\bar{x}'_2,t_{0-};2')
\label{sol1}
\eea
is clearly the solution of the homogeneous MSH with the correct 
boundary conditions, see Eq. (\ref{bcfg2}). We can manipulate Eq. 
(\ref{sol1})  by introducing a linear combination of $\d$-functions on the 
contour
\be
\d_{-}(z)\equiv\d(z,t_{0-})-\d(z,t_{0+}).
\ee
The spectral function appearing in Eq. (\ref{sol1}) can be written as 
\be
A(1;x'_1,t_{0-})=i\int_\gamma d\bar{z}\;g(1;x'_1,\bar{z})\delta_-(\bar{z})
\ee
and
\be
A(x_1,t_{0-};1')=-i\int_\gamma 
d\bar{z}\;\delta_-(\bar{z})g(x_1,\bar{z};1').
\ee
Inserting these expressions into Eq. (\ref{sol1}) we find
\be
\tilde{g}_2(1,2;1',2')=\int d\bar{1}d\bar{2}d\bar{1}'d\bar{2}'
g(1;\bar{1})g(2;\bar{2})C_2(\bar{1},\bar{2};\bar{1}',\bar{2}')
g(\bar{1}';1')g(\bar{2}';2')
\ee
where
\be
C_2(\bar{1},\bar{2};\bar{1}',\bar{2}')=
\delta_-(\bar{z}_1)\delta_-(\bar{z}_2)
C_2(\bar{x}_1,\bar{x}_2;\bar{x}'_1,\bar{x}'_2)
\delta_-(\bar{z}'_1)\delta_-(\bar{z}'_2)
\ee
is the two-particle initial-correlation function. In conclusion 
$\tilde{g}_{2}$ can be written in terms of $g$ and $\G_{n}$ with 
$n\leq 2$. Since $g_{2}=|g|_{2}+\tilde{g}_{2}$ this result provides a 
decomposition of $g_{2}$ in terms of $g$ and reduced $n$-particle 
denity matrices.

The generalization of Wick's theorem to $g_{n}$ reads
\be
g_n=|g|_n+\;
\sum_{l=1}^{n-2}\sum_{PQ}
(\pm)^{|P+Q|}\;|g|_l(P;Q')\;\tilde{g}_{n-l}(\breve{P};\breve{Q}')
\;+\;\tilde{g}_n.
\label{gen-wick0}
\ee
In this formula $P$ and $Q$ are a subset of $l$ ordered indices 
between $1$ and $n$ whereas $\breve{P}$ and $\breve{Q}$ is the 
ordered complementary subset. For instance if $n=3$ and $l=1$ then we 
can have $P=1$ and hence $\breve{P}=(2,3)$, or $P=2$ and hence 
$\breve{P}=(1,3)$, or $P=3$ and hence $\breve{P}=(1,2)$. The sign of the 
various terms is given by $|P+Q|=\sum_{i=1}^{l}(p_{i}+q_{i})$ where 
$p_{i}$ and $q_{i}$ are the indices in the $l$-tuple $P$ and $Q$.  
The solution of the homogeneous MSH with the correct boundary 
conditions is
\be
\tilde{g}_{n}(1\ldots n;1'\ldots n')=\int
g(1;\bar{1})\ldots g(n;\bar{n})C_{n}(1\ldots n;1'\ldots 
n')g(\bar{n}';n')\ldots g(\bar{1}';1')
\label{cn-block}
\ee
where the integral is over all barred variables and the $n$-particle 
initial-correlation functions are given by
\be
C_{n}(1\ldots n;1'\ldots n')=\d_{-}(z_{1})\ldots\d_{-}(z_{n})
C_{n}(x_{1}\ldots x_{n};x'_{1}\ldots 
x'_{n})\d_{-}(z'_{1})\ldots\d_{-}(z'_{n})
\ee
with
\be
C_{n}=\G_{n}-\sum_{l=1}^{n-2}\sum_{PQ}(\pm)^{|P+Q|}
|\G|_{l}(X_{P};X'_{Q})C_{n-l}(X_{\breve{P}};X_{\breve{Q}}')
-|\G|_{n}.
\ee
This is a recursive formula for the $C_{n}$. The collective 
coordinate $X_{P}=(x_{p_{1}}\ldots x_{p_{l}})$ is a subset of the
coordinates $(x_{1}\ldots x_{n})$ and similarly the collective 
coordinate $X'_{Q}=(x'_{q_{1}}\ldots x'_{q_{l}})$ is a subset of the
coordinates $(x'_{1}\ldots x'_{n})$. 

We defer the reader to Ref. 
\cite{vls.2012} for the proof of the generalized Wick theorem. Here 
we observe that with the generalized Wick theorem we can express 
$g_{n}$ in terms of $g$ and $\G_{n}$. Since $\G_{n}$ can 
easily be calculated from $\hat{\r}$ the generalized 
Wick theorem is especially suited to do MBPT when we know  
$\hat{\r}$ instead of $\hat{H}^{\rm M}$. We further observe that the 
generalized Wick theorem has the same mathematical structure of the 
Laplace expansion for the permanent/determinant of the sum of two 
matrices $A$ and $B$ \cite{svlbook}
\be
|A+B|_{m} = |A|_{m} + \sum_{l=1}^{m-1} \sum_{PQ} 
(\pm)^{|P+Q|}
|A|_{l}(P;Q) |B|_{m-l} (\breve{P};\breve{Q})  
+  |B|_{m}  
\label{sumdet}
\ee
where $|A|_{l}(P;Q)$ is the permanent/determinant of the $l\times 
l$ matrix obtained with the rows $P$ and the 
columns $Q$ of the matrix $A$. The same notation has been used for 
the matrix $B$.
With the identification $A_{kj}=g(k;j')$ and $|B|_{m-l} 
(\breve{P};\breve{Q}) = \tilde{g}_{m-l} 
(\breve{P};\breve{Q}')$   
for $l=1 \ldots m-2$ and the definition $\tilde{g}_1 \equiv 0$
Eqs. (\ref{gen-wick0}) and (\ref{sumdet})  become identical.
We can thus symbolically write the 
generalized Wick theorem  as 
\be
g_{m} = |g + \tilde{g}|_{m}
\label{gen-wick2}
\ee
whose precise meaning  is given by Eq. (\ref{gen-wick0}).

\section{Relation with the Konstantinov-Perel' formalism}

\begin{figure}[h]
\includegraphics[width=18pc]{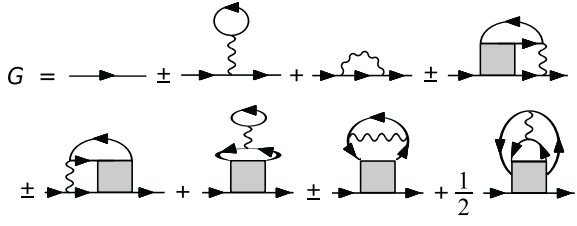}\hspace{4pc}%
\begin{minipage}[b]{16pc}
    \caption{\label{GFexp} Diagrammatic expansion of the Green's function to
     first order in the interaction.}
\end{minipage}
\end{figure}

In Eq.(\ref{genexp}) we have seen how we can expand the Green's 
function $G$ into noninteracting Green's functions $g_n$ satisfying the generalized
Wick theorem (\ref{gen-wick0}). This can be used to define a diagrammatic expansion
of the Green's function. Let us see what terms we get when we insert Eq.(\ref{gen-wick0})
into Eq.(\ref{genexp}).
The first term $|g|_n$ in Eq.(\ref{gen-wick0}) generates the usual series of
connected diagrams for the Green's function in powers of the interaction
(the disconnected diagrams are zero since they are integrated from $t_0$ to $t_0$
and have no external points). The remaining terms in Eq.(\ref{gen-wick0})
are linear in the functions $\tilde{g}_m$ with $m=2,\ldots,n$. As a consequence of Eq.(\ref{cn-block}) these
functions can be diagrammatically represented by blocks $C_m$ with $m$ ingoing and $m$ outgoing $g$ lines. Since the Wick expression (\ref{gen-wick0})
is linear in the $\tilde{g}_m$ each diagram contribution to the Green's function contains at
most one correlation block. In Fig. \ref{GFexp} we display the diagrams for the Green's function up to first order in the
interaction involving four diagrams with a $C_2$ block and one diagram with a $C_3$ block.
The last three diagrams, however, vanish since for those diagrams there are internal time-integrations for
Green's funtion lines that enter and leave a $C_m$ block that can be reduced to a point, since the initial correlation block
only exists at time $t_{0-}$.
The fact that the $C_m$ blocks are not repeated within a single diagram prevents us from deriving an irreducible self-energy in terms of them.
We can, however, define a reducible self-energy $\sigma_r$ and write the Green's function as
\be
G(1;2) = g(1;2) + \int_{\bar{\gamma}} d\bar{1}d\bar{2} \, g(1;\bar{1}) \sigma_r (\bar{1};\bar{2}) g(\bar{2};2) 
\label{gred}
\ee
where we used the notation $\bar{\gamma}$ for the contour of Fig.\ref{basiccontour} to distinguish it from the
extended contour $\gamma$ that we will use later.
The reducible self-energy can be split into three contributions. 
These are the sets of self-energy diagrams that start with an interaction line at their entrance vertices,
and end with a correlation block at their exit vertices, denoted by $\sigma_r^{L}$, the diagrams that start with a correlation block and 
end with an interaction line, denoted by $\sigma_r^R$, and the remaining diagrams (which either contain no correlation block or contain a correlation block only attached
to internal vertices) which we will denoted by $\tilde{\sigma}_r$. 
The labels $L$ and $R$ therefore refer to the location of an interaction line at the entrance vertices.
We can thus write
\be
\sigma_r = \tilde{\sigma}_r + \sigma_r^L + \sigma_r^R
\ee
where
\bea
\sigma_r^L (1;2) &=&  \sigma_r^L (1;x_2) \delta_{-} (z_2) \\
\sigma_r^R (1;2) &=& \delta_{-} (z_1) \sigma_r^R (x_1;2) .
\eea
We emphasize that $\sigma_r$ is a reducible self-energy and therefore 
it should not be confused with the self-energy of Ref. \cite{h.1975}.
We will now proceed to connect the reducible self-energy $\sigma_r$ to the irreducible self-energy appearing in the 
equation of motion for the Green's function on the extended contour. As discussed in the introduction
the initial ensemble is of the form $\hat{\rho} = e^{-\hat{X}}/ \mathrm{Tr} [ e^{-\hat{X}}]$, where $\hat{X}= \sum_m \hat{X}_m$ is in general
a sum of $m$-body operators of the form
\be
\hat{X}_m = \frac{1}{m!} \int dx_1 \ldots dx_m'  \, v_m (x_1, \ldots , x_m, x_1', \ldots, x_m') 
\hat{\psi}^\dagger (x_1) \ldots \hat{\psi}^\dagger (x_m) \hat{\psi} (x_m') \ldots \hat{\psi} (x_1')
\ee
The functions $v_m$ must be chosen in such a way that the pre-scribed 
density matrices of Eq.(\ref{gamman}) are obtained. 
This is, in general, a difficult task. If we, however, assume that we 
have succeeded in this task we can define the 
Matsubara Hamiltonian $\hat{H}^{\rm{M}}=\hat{X}/\beta$ and use the 
contour of Fig. (\ref{exactcontour}) and the expression for 
the Green's function (\ref{gn3}) to expand the Green's function 
diagrammatically in powers of $G_0$ using the standard Wick theorem 
of Eq.(\ref{zeroGn}) \cite{d.1984,w.1991,MozorovRoepke}. The diagrammatic rules for the Green's function 
in the case of $m$-body operators are given in Ref. \cite{w.1991}. 
If we collect the irreducible pieces of this expansion in an 
irreducible self-energy $\Sigma$ then we obtain a Dyson equation on 
the extended contour
\be
G(1;2) = G_0 (1;2) + \int_\gamma d3 d4 \, G_0 (1;3) \Sigma (3;4) G 
(4,2).
\label{Dyson}
\ee
It will be more convenient to write this as
equations of motion
\bea
(i  \partial_{z_1} - h(1)) G(1;2) &=& \delta (1;2) + \int_\gamma d3 
\, \Sigma (1;3) G (3;2) 
\label{KB1}\\
G(1;2) ( -i \overleftarrow{ \partial_{z_2}} - h(2) ) &=& \delta (1;2) 
+ \int_\gamma d3 \, G (1;3) \Sigma (3;2)
\label{KB2}
\eea
When split into various components these are simply the Kadanoff-Baym equations on the extended contour.
Let us now write the contour $\gamma$ as $\gamma = \bar{\gamma} \oplus \gamma^{\rm{M}}$, where $\gamma^{\rm{M}}$
denotes the vertical (or Matsubara) track of the contour and $\bar{\gamma}$ the remaining piece, which is 
identical to the contour of Fig. \ref{basiccontour}. 
The last term on the r.h.s of Eq.(\ref{KB1}) can then be written as
\be
\int_{\gamma} d3 \, \Sigma (1;3) G (3;2) = \int_{\bar{\gamma}} d3 \, 
\Sigma (1;3) G (3;2)
- i \int dx_3 \int_0^\beta d\tau \, \Sigma^\rceil (1;x_3 \tau) 
G^\lceil (x_3 \tau;2)
\label{sig_G}
\ee
where we introduced the parametrization $z=t_0-i\tau$ on the vertical track $\gamma^{\rm{M}}$. For a general
function $A(z,z')$ on the contour (spatial coordinates suppressed) we further defined
\bea
A^\rceil (z=t_{\pm},\tau) &=& A (z=t_{\pm},t_0-i \tau) \\
A^\lceil (\tau, z=t_{\pm}) &=& A (t_0-i \tau, z=t_{\pm}) .
\eea
From the Dyson equation (\ref{Dyson}) and the Langreth rules on the 
contour $\gamma$ we can further derive that \cite{svlbook,sa.2004}
\be
G^\lceil (1;2) = - i \int d\bar{x} \, G^{\rm{M}} (1;\bar{x} t_0) 
G^{\rm{A}} (\bar{x} t_0; 2)
+ [ G^{\rm{M}} \star \Sigma^\lceil \cdot G^{\rm{A}} ] (1;2)
\label{Gleft}
\ee
where $\star$ denotes a convolution between $t_0$ and $t_0-i\beta$ on the vertical track and
$\cdot$ denotes a convolution between $t_0$ and $\infty$. 
We further defined the advanced and Matsubara Green's
functions as
\bea
G^{\rm{A}} (1;2) &=& - \theta (t_2- t_1) [ G^> (1;2) - G^< (1;2)] \\
G^{\rm{M}} (1;2) &=& G (x_1 t_0-i \tau_1; x_2 t_0-i \tau_2)
\eea
If we further use that
\be
\int_{\bar{\gamma}} dz \, \delta_{-} (z) G(\bar{x} z; x t_{\pm} ) = 
G^{\rm{A}} (\bar{x} t_0; x t )
\ee
we find by inserting Eq.(\ref{Gleft}) into the last term of Eq.(\ref{sig_G}) that
\bea
\lefteqn{- i \int dx_3 \int_0^\beta d\tau \, \Sigma^\rceil (1;x_3 
\tau) G^\lceil (x_3 \tau;2) = }\nonumber \\
&& \int_{\bar{\gamma}} d3 \, [ \Sigma^\rceil \star G^{\rm{M}}Ê\star 
\Sigma^\lceil ] (1;3) G(3;2) 
-i \int_{\bar{\gamma}} d3 \, [ \Sigma^\rceil \star G^{\rm{M}} ] 
(1;x_3 t_0) \delta_{-} (z_3) G(3;2)
\eea
If we therefore define $\Sigma_{\rm{L}}$ by
\be
\Sigma_{\rm{L}} (1;2) = -i [ \Sigma^\rceil \star G^{\rm{M}} ] (1;x_2 t_0) \delta_{-} (z_2) ,
\ee
we can rewrite the equation of motion (\ref{KB1}) for the Green's function as
\bea
(i  \partial_{z_1} - h(1)) G(1;2) = \delta (1;2)  
+ \int_{\bar{\gamma}} d3 \, [ \Sigma + \Sigma^\rceil \star G^{\rm{M}}Ê
\star \Sigma^\lceil + \Sigma_{\rm{L}} ] (1;3)G (3;2) .
\label{KB1_2}
\eea
A similar procedure can be carried out for the adjoint equation (\ref{KB2}). We find
\bea
G(1;2) ( -i \overleftarrow{ \partial_{z_2}} - h(2) ) &=& \delta (1;2) + 
\int_{\bar{\gamma}} d3 \, G (1;3) [ \Sigma + \Sigma^\rceil \star 
G^{\rm{M}}Ê\star \Sigma^\lceil + \Sigma_{\rm{R}} ] (3;2)
\label{KB2_2}
\eea
where we defined
\be
\Sigma_{\rm{R}} (1;2) = -i \delta_{-} (z_1)  [  G^{\rm{M}} \star 
\Sigma^\lceil] (x_1 t_0;2 ) .
\ee
Given the self-energy and the Green's function with arguments on the imaginary track $\gamma^{\rm{M}}$, we can regard
Eqs. (\ref{KB1_2}) and (\ref{KB2_2}) as equations of motion for the Green's function $G$ on the contour $\bar{\gamma}$.
These equations can be integrated using the noninteracting Green's function $g$ of Eq.(\ref{g_0}) since it satisfies
\be
(\pm i)\lim_{z_{1},z'_{1}\ra 
t_{0-}}g(1;1')= (\pm i)\lim_{z_{1},z'_{1}\ra 
t_{0-}} G(1;1') = \G(x_{1};x'_{1}).
\label{g_0_1}
\ee
If we, therefore, define the total self-energy as
\be
\Sigma_{\rm{tot}} = \Sigma + \Sigma^\rceil \star G^{\rm{M}}Ê\star \Sigma^\lceil + \Sigma_{\rm{L}} + \Sigma_{\rm{R}}
\ee
we can write $G$ in terms of two equivalent Dyson equations
\bea
G(1;2) &=& g (1;2) + \int_{\bar{\gamma}} d3 d4 \, g(1;3) 
\Sigma_{\rm{tot}} (3;4) G(4;2) 
\label{D1} \\
 G(1;2)  &=& g(1;2) + \int_{\bar{\gamma}} d3 d4 \, G(1;3) 
 \Sigma_{\rm{tot}} (3;4) g(4;2) 
\label{D2}
\eea
To check that these equations are equivalent to the Eqs. 
(\ref{KB1_2}) and (\ref{KB2_2}) we need to be careful.
The standard approach is to act with the operator of the form $i 
\partial_z - h$ and its adjoint on both Dyson equations
and use the equation of motion for $g$ 
\bea
(i \partial_{z_1}  - h(1) ) g(1;2) &=& \delta (1;2) 
\label{g_mot1} \\
g(1;2) (-i \overleftarrow{\partial_{z_2}}- h(2)) &=& \delta (1;2)
\label{g_mot2}
\eea
We need to be careful, however, since we cannot change integration and differentiation in the presence of delta-functions under the integral sign.
The relevant integrals over the delta-functions need to be done first before we use Eqs.(\ref{g_mot1}) and (\ref{g_mot2}).
In Eq.(\ref{D1}) we have an integral of the form
\be
 \int_{\bar{\gamma}} d3 \, g(1;3) \Sigma_{\rm{R}} (3;4) = -i 
 \int dx_3 [g^> (1;x_3 t_0) - g^< (1;x_3 t_0)]  [  G^{\rm{M}} \star 
 \Sigma^\lceil] (x_3 t_0;4 ) 
\ee
On the right hand side of this equation we recognize the contour spectral function of Eq.(\ref{spectral}) that satisfies 
Eq. (\ref{spectral_mot}). We therefore see that
\be
(i \partial_{z_1}  - h(1) ) \int_{\bar{\gamma}} d3 \, g(1;3) 
\Sigma_{\rm{R}} (3;4) = 0
\ee
Similarly we have
\be
 \left( \int_{\bar{\gamma}} d3 \, \Sigma_{\rm{L}} (3;4) g (4;2) \right) (-i \overleftarrow{\partial_{z_2}}- h(2)) = 0
\ee
Then by acting with $i \partial_{z_1}  - h(1) $ on Eq.(\ref{D1}) we see that we recover Eq.(\ref{KB1_2}). Similarly by
acting with $-i \overleftarrow{\partial_{z_2}}- h(2)$ from the left on Eq.(\ref{D2}) we recover Eq.(\ref{KB2_2}). 
It only remains to check that the Dyson Eqs.(\ref{D1}) and (\ref{D2}) satisfy the correct boundary conditions.
Since in the limit $z_1,z_2 \ra t_{0-}$ the contribution for the integrals on the r.h.s. of the equations vanish we 
see that the condition (\ref{g_0_1}) is indeed satisfied. 

Now we are ready to discuss the connection between the formulation based on the initial correlation blocks and the
formalism based on integrations along the imaginary track. By comparing Eq.(\ref{D1}) to Eq.(\ref{gred}) we see that
\be
\sigma_r = \Sigma_{\rm{tot}} + \Sigma_{\rm{tot}} g \Sigma_{\rm{tot}} + \Sigma_{\rm{tot}} g \Sigma_{\rm{tot}} g \Sigma_{\rm{tot}} + \ldots =
\Sigma_{\rm{tot}} \frac{1}{1-g \Sigma_{\rm{tot}}}
\ee
and hence
\be
\Sigma_{\rm{tot}} = \sigma_r (1 - g \Sigma_{\rm{tot}}) = \sigma_r - \sigma_r g \sigma_r + \sigma_r g \sigma_r g \sigma_r - \ldots = \sigma_r
\frac{1}{1 + g \sigma_r}
\ee
This yields the expansion of the irreducible self-energy 
$\Sigma_{\rm{tot}}$ in terms of the Green's functions $g$ and the 
correlation blocks $C_m$. We would like to mention that 
$\Sigma_{\rm{tot}}$ could in principle be calculated from the 
appropriate extention of the Hedin equations to include initial 
correlations \cite{Bonitz:PRE95}. This would lead to an expansion of 
$\S_{\rm tot}$ in terms of the dressed Green's function $G$ and 
correlation blocks. However, the iterative solution of these equations 
depend on the starting point. In particular if we start with a 
self-energy which contains only a $C_{2}$-block then the iterative 
procedure cannot generate diagrams with $C$-blocks of higher order.

\section{Conclusions}

We presented a unified framework for equilibrium and nonequilibrium many-body perturbation theory.
The most general formalism for nonequilibrium many-body theory for general initial states is based on 
the Keldysh contour to which we attach a vertical track describing a general initial state. This idea goes
back to the works of Konstantinov and Perel' (who considered equilibrium initial states), Danielewicz and Wagner.
On this contour we can straightforwardly prove a Wick theorem by solving the noninteracting Martin-Schwinger
hierarchy for the noninteracting many-body Green's functions with KMS boundary conditions. This short
proof of Wick's theorem does not need any of the usually introduced theoretical concepts such as normal
ordering and contractions. The statement is simply that the noninteracting $m$-particle Green's function is
a determinant or permanent of one-particle Green's functions.
We showed how the various other well-known formalisms of Keldysh, Matsubara and
the zero-temperature formalism can be derived as special cases that arise under different assumptions.
We further discussed a generalized Wick theorem for general initial states on the Keldysh contour. 
It again arises as a solution of the noninteracting Martin-Schwinger
hierarchy for the noninteracting many-body Green's functions but this time with initial conditions
specified by initial $m$-body density matrices. The final result of Eq.(\ref{gen-wick2}) is an elegant alternative
to the Wick theorem of Eq.(\ref{zeroGn}) for KMS boundary conditions.  
We finally showed how the formalisms based on the Keldysh and Konstantinov-Perel'-contours are related for the
case of general initial states.

\end{document}